\documentclass[aps,pra,twocolumn]{revtex4}
\usepackage{epsfig,dsfont,amssymb,amsmath,amsthm,amsfonts,amsbsy,mathrsfs}
\usepackage{graphicx}
\usepackage{bm}%bold math
\begin{document}

\title{Continuous-mode effects and photon-photon phase gate performance}

\author{Bing He} 
\author{Artur Scherer} 
\affiliation{Institute for Quantum Information Science, University of Calgary, Alberta, 
Canada T2N 1N4}

\begin{abstract}
The effects arising from the inherent continuous-mode nature of photonic pulses
were poorly understood but significantly influence the performance of quantum devices employing photonic pulse interaction in nonlinear media. Such effects include the entanglement between the continuous wave-vector modes due to pulse interaction as well as the consequence of a finite system bandwidth. We present the first analysis on these effects for interactions between single-photon pulses, demonstrating their impact on the performance of quantum phase gates based on such process. Our study clarifies a realistic picture of this type of quantum 
devices.
\end{abstract}
\maketitle

\section{Introduction}
Deterministic photon-photon phase gate is a key building block to construct circuits for the scalable all-optical quantum information processing. Cross-phase modulation (XPM) between slow pulses in media under electroma\-gnetically induced transparency (EIT) conditions \cite{s-i-96,f-i-05} (or with similar properties) is the main route towards such a gate. Considerable theoretical developments  (see, e.g., \cite{h-h-99, l-i-00, p-k-02, m-f-04, p-m-04, a-05, f-k-05, r-v-04, w-s-06, m-10, bhe-11-1, headon-2, bhe-11-2, bec-11}) as well as experimental studies (see, e.g., \cite{ex1, ex2, ex3, ex4, ex5, ex6, ex7}) have been undertaken to explore the EIT-based XPM and alternative approaches aiming at realizing photonic two-qubit gates. 

In optics-based quantum computing, a conditional phase of $\pi$ radians needs to be implemented by photon-photon phase gates. For achieving such a large phase there are the proposals \cite{l-i-00, p-k-02, p-m-04, r-v-04, w-s-06} of making pulses co-propagate, so that the generic weak pulse interaction could be compensated by the prolonged interaction time. Another requirement for an ideal phase gate is the uniformness of conditional phase, as given by the mapping $|1\rangle_1|1\rangle_2\rightarrow e^{i\theta}|1\rangle_1|1\rangle_2$, i.e., the same phase $\theta$ is induced 
for each mode $k$ of a continuous-mode photon in the state $|1\rangle=\int_{-\infty}^{\infty} dk~ \xi(k)\hat{a}^{\dagger}(k)|0\rangle$, where $\xi(k)$ is the pulse profile in wave-vector space. The currently dominant understanding is that a homogeneous phase could be possible if the interaction between two pulses is averaged out by letting one pulse completely go through the other \cite{m-f-04, a-05, f-k-05, m-10, headon-2}. 

Since a photon is not a point particle, photon-photon interactions in nonlinear media should be modeled as interacting quantum fields of continuous modes. Using this picture, we show that the above-mentioned notions
are generally invalid. The essential effects in realistic single-photon XPM are clarified here for the first time.

\section{continuous-mode effects in photon-photon interactions}

We first provide a theoretical framework for the XPM between photons. The interaction between two pulses in a medium of any type of atomic structure realizing EIT can be translated 
into that between two dark-state polariton fields $\hat{\Psi}_l(z,t)=\frac{1}{\sqrt{2\pi}}\int_{-\infty}^{\infty} \hat{a}_{l}(k) e^{ikz}dk$ ($l=1,2$) with $[\hat{a}_{i}(k),\hat{a}^{\dagger}_{j}(k')]=\delta_{i,j}\delta(k-k')$ \cite{ex}. By neglecting the pulse loss and deformation, as well as the possible self-phase modulation term which has no effect on photon-photon interactions \cite{bhe-11-1, headon-2, bhe-11-2}, one has the following equations of motion for the slowly varying and transversely well confined polariton fields \cite{f-k-05, headon-2, bhe-11-1, bhe-11-2}:
\begin{equation}
\left(\partial_t+v_i\partial_{z}\right)\hat{\Psi}_{i}(z,t)=
-i\hat{\alpha}_i(z,t)\hat{\Psi}_{i}(z,t) \,
\label{motion}
\vspace{-0.1cm}
\end{equation}
where $v_i$ are the pulse group velocities. The term $\hat{\alpha}_i(z,t)=\int dz'\Delta(z-z')\hat{\Psi}_{3-i}^{\dagger}(z',t)\hat{\Psi}_{3-i}(z',t)$ could come from a general interaction potential $\Delta(z-z')$. The pulse interaction in the experimentally studied XPM thus far \cite{ex1,ex2,ex3,ex4, ex5, ex6, ex7}, for instance, can be modeled by a contact potential $\Delta(z-z')=\chi \delta(z-z')$, where 
$\chi$ approximated by a real quantity is the nonlinear rate determined by the specific system parameters. The potential $\Delta(z-z')$ considered here acts instantaneously; see \cite{s-06} for a study on the non-instantaneous effects. From the field-theoretic viewpoint, Eq. (\ref{motion}) is obtained by the equation of motion $i\hbar\partial_t\hat{\Psi}_i=\delta\hat{H}/\delta\hat{\Psi}^{\dagger}_i$ for non-relativistic fields, where the Hamiltonian $\hat{H}=\hat{K}+\hat{V}$ consists of the kinetic term $\hat{K}=\sum_{l=1}^{2}\int dz  v_l\hat{\Psi}_l^{\dagger}(z)\frac{\hbar}{i}\nabla_{z}\hat{\Psi}_l(z)$ and the interaction term $\hat{V}=\hbar\int dz\int dz'\hat{\Psi}_1^{\dagger}(z)\hat{\Psi}_2^{\dagger}(z')\Delta(z-z')
\hat{\Psi}_2(z')\hat{\Psi}_1(z)$. 

The pulse interaction would evolve the input state to $\hat{U}(t)|1\rangle_1 |1\rangle_2=\int dk\int dk'\zeta(k,k',t)\hat{a}^{\dagger}(k)\hat{b}^{\dagger}(k')|0\rangle$, 
where $\hat{U}(t)=\mathbb{T} \exp\{-i\int_0^t dt'\hat{H}(t')\}$ ($\mathbb{T}$ denotes the time-ordering operation and $\hbar\equiv 1$ is adopted hereafter). It is convenient to use the 
{\it two-particle function}, $\psi(z_1,z_2,t)=\langle 0|\hat{\Psi}_1(z_1,t)\hat{\Psi}_2(z_2,t)|\Phi_{\mbox{\scriptsize in}}\rangle=\frac{1}{2\pi}\int dk\int dk'\zeta(k,k',t)e^{i k z_1}e^{ik' z_2}$, to study the evolution of the initial state $|\Phi_{\mbox{\scriptsize in}}\rangle=|1\rangle_1|1\rangle_2$. 
 To indicate how close a realistic XPM is to the ideal one $|1\rangle_1|1\rangle_2\rightarrow e^{i\theta}|1\rangle_1|1\rangle_2$ for phase gates, we use the {\it fidelity} $F$.
As figures of merit to characterize a realistic XPM, the fidelity $F$ and conditional phase $\theta$ are determined by the overlap \cite{bhe-11-1,bhe-11-2} 
\begin{eqnarray}
&\sqrt{F}e^{i\theta}=
\langle \Phi_0 | \Phi_{\mbox{\scriptsize out}}\rangle =\int\! dz_1\!\int \!dz_2 \psi^{\ast}_0(z_1,z_2, t)
\psi(z_1,z_2, t),& \nonumber \\&&
\label{over-lap}
\end{eqnarray}
where $\psi_{0}(z_1,z_2, t)$ is the two-particle function from the field operator
equations $(\partial_t+v_i\partial_{z})\hat{\Psi}_{i}(z,t)=0$, corresponding to the freely evolved state $|\Phi_{0}\rangle=e^{-i\hat{K}t}|\Phi_{\mbox{\scriptsize in}}\rangle$. A similar formula in the discrete 
form is given in \cite{bana-09}.

The output state of a realistic XPM, $|\Phi_{\mbox{\scriptsize out}}\rangle=\int dk\int dk'\zeta(k,k',t)\hat{a}^{\dagger}(k)\hat{b}^{\dagger}(k')|0\rangle$, is generally entangled between the wave-vector modes $k$ and $k'$ of the individual photons. Such field modes entanglement was widely neglected in the previous researches on photon-photon gates, though its effect on XPM was conjectured in \cite{f-i-05}. It is conceivable that the entanglement would lower the fidelity $F$. Yet, for clarifying the issue, a relation between the amount of entanglement generated in pulse interaction and the corresponding gate operation fidelity has to be found. Here we quantify the field mode entanglement with the linear entropy $S_L=1-{\mbox Tr}\rho_i^2$ ($\rho_i$ are the reduced density matrices of the bipartite state $|\Phi_{\mbox{\scriptsize out}}\rangle$), which takes the following closed form:
\begin{eqnarray}
S_L(t)&=& 1-\int dz_1dz_2dz_3dz_4\Big\{\psi(z_1,z_2,t)
\psi^{\ast}(z_3,z_2,t)\nonumber\\
&\quad\times& \psi(z_3,z_4,t)\psi^{\ast}(z_1,z_4,t)\Big\};
\label{entropy}
\end{eqnarray}
see Appendix for the proof.
Compared with the Schmidt decomposition method to characterize bi-photon entanglement \cite{spdc, l-e-04}, this formula provides an exact measure of such entanglement.
 
In reality the medium for pulse interaction carries a finite bandwidth $\Delta\omega_s$, which is connected to the width of EIT transparency window \cite{l-i-00,p-m-04}. Hence, the equal-time commutator for the field operator $\hat{\Psi}_l(z,t)=\frac{1}{\sqrt{2\pi}}\int_{-k_s}^{k_s} \hat{a}_{l}(k) e^{ikz}dk$ after being imposed a cut-off by the bound of the wave-vector mode $k_s=\Delta \omega_s/(2c)$ becomes 
\begin{eqnarray}
[\hat{\Psi}_{i}(z_1),\hat{\Psi}^{\dagger}_{j}(z_2)]&=&\delta_{ij}(k_s/\pi) \mbox{sinc}\big (k_s(z_1-z_2)\big)\nonumber\\
&\equiv &\delta_{ij}C(z_1-z_2), 
\label{com}
\end{eqnarray}
where $\mbox{sinc}(x)\equiv\sin (x)/x$. Substituting the formal solution 
of (\ref{motion}), $\hat{\Psi}_{i}(z,t)=\exp\{-i\int_0^{t}dt'\hat{\alpha}_{i}\big(z- v_{i}(t-t')\big)\}\hat{\Psi}_{i}(z- v_it,0)$, where $\hat{\alpha}_{i}(z)$ originates from the contact potential $\chi \delta(z-z')$, into $\psi(z_1,z_2,t)=\langle 0|\hat{\Psi}_1(z_1,t)\hat{\Psi}_2(z_2,t)|\Phi_{\mbox{\scriptsize in}}\rangle$, while considering the commutator in (\ref{com}), yields the general two-particle function
\begin{eqnarray}
&& \psi (z_1,z_2,t)= \langle 0|\hat{\Psi}_{1}(z'_1)\hat{\Psi}_{2}(z'_2)|\Phi_{\mbox{\scriptsize in}}\rangle
+ \langle 0|\sum_{n=1}^{\infty}\frac{(i\chi)^n}{n!}
 \nonumber\\
&&\times \int_0^t dt_{n} \cdots \int_0^t dt_1  C(v_rt_{n-1}-v_rt_n)
\cdots   C(v_rt_1-v_rt_2)\nonumber\\
&& \times  C(z'_2-z'_1-v_r t_1) \hat{\Psi}_{1}(z'_2-v_rt_n)\hat{\Psi}_2(z'_2)|\Phi_{\mbox{\scriptsize in}}\rangle
\label{2-p}
\end{eqnarray}
at the time $t$, where $z'_i=z_i-v_it$ and $v_r=v_1-v_2$. 
The second term on the right side of Eq.~(\ref{2-p}) arises from photon-photon interaction. The norm of the two-particle function $\psi (z_1,z_2,t)$ is not preserved as a consequence of the deviation of the field operator commutator $C(z-z')$ from the delta function $\delta(z-z')$, so the output two-photon function $\psi (z_1,z_2,t)$ has to be normalized prior to 
calculating the fidelity and linear entropy using Eqs. (\ref{over-lap}) and (\ref{entropy}). The system bandwidth $\Delta\omega_s$ effectively results in a non-unitary evolution, though the wave-vector modes $k$, $k'$ of two photons are still continuous after imposing the cut-off $k_s$. This could be understood by the restriction $-k_s\leq k,k'\leq k_s$ on the matrix elements $\langle k,k'|\hat{U}(t)|k,k'\rangle$ of the evolution operator $\hat{U}(t)$, causing the loss of the orthonormal relation for the matrix elements $\langle k,k'|\hat{U}(t)|k,k'\rangle$; c.f. the inequality $\int_{-k_s}^{k_s} dl \int_{-k_s}^{k_s} dl' \langle k,k'|\hat{U}|l,l'\rangle\langle l,l'|\hat{U}^{\dagger}|q,q'\rangle\neq \int_{-\infty}^{\infty} dl \int_{-\infty}^{\infty} dl' \langle k,k'|\hat{U}|l,l'\rangle\langle l,l'|\hat{U}^{\dagger}|q,q'\rangle=\delta(k-q)\delta (k'-q')$. 

\section{impact on photonic phase gate performance}
An important situation we analyze here is the XPM between two co-propagating pulses \cite{l-i-00, p-k-02, p-m-04, r-v-04, w-s-06}. With the relative velocity $v_r=0$, the two-particle function in Eq. (\ref{2-p}) reduces to (see also Ref. \cite{p-m-04})
\begin{eqnarray}
&&\psi(z_1,z_2,t)=f_1(z_1-vt)f_2(z_2-vt)
+f_1(z_2-vt)\nonumber\\
&&\times f_2(z_2-vt) \mbox{sinc}\big (\frac{\Delta \omega_{s}}{2c}(z_1-z_2)\big )(e^{i\Phi}-1),
\label{2-p-e}
\end{eqnarray}
where $f_i(z)=\langle 0|\hat{\Psi}_i(z)|1\rangle$, $\Phi= \chi \Delta \omega_s t /(2\pi c)$ and $v_1=v_2=v$. 
Using the normalized form of this two-particle function, one will obtain from Eq. (\ref{over-lap}) the following relations to determine the conditional phase and fidelity: 
\begin{equation}
\vspace{-0.1cm}
\tan\theta=\frac{C_1\sin\Phi}{1-C_1+C_1\cos\Phi},
\label{c-phase}
\vspace{-0cm}
\end{equation}
\begin{equation}
\vspace{-0.1cm}
F=\frac{1-4C_1(1-C_1)\sin^2\frac{\Phi}{2}}{1-4(C_1-C_2)\sin^2\frac{\Phi}{2}}, 
\label{fidelity}
\vspace{-0cm}
\end{equation}
where $C_1=\int dZ_1\int dZ_2 f^{\ast}_1(Z_1)f_1(Z_2)|f_2(Z_2)|^2\mbox{sinc}[k_0(Z_1-Z_2)]$, $C_2=\int dZ_1\int dZ_2 |f_1(Z_2)|^2|f_2(Z_2)|^2 (\mbox{sinc}[k_0(Z_1-Z_2)])^2$. The dimensionless variables of the integrals are $Z_i=(z_i-vt)/\sigma$, where $\sigma$ is the pulse size in medium; the system parameter is defined as $k_0=v/(2c)\Delta \omega_{s}/\Delta \omega_{p}$, in proportion to the ratio of the system bandwidth $\Delta \omega_{s}$ to the pulse bandwidth $\Delta \omega_{p}$ (the reciprocal of the pulse duration). 
The factor $v/(2c)$ in the parameter $k_0$ reflects the pulse compression in EIT media.

\begin{figure}[b!]
\vspace{-0cm}
\centering
\epsfig{file=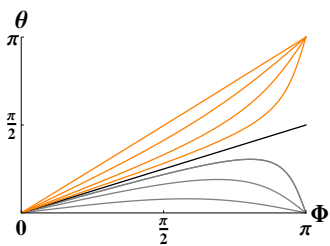,width=0.5\linewidth,clip=} 
{\vspace{-0cm}\caption{\label{Fig:fhi_vs_Phi} (color online) Relation between the conditional phase $\theta$ and the XPM phase $\Phi$ for various $C_1$ values from the lower to the upper: $C_1=0.20$, $0.36$, $0.45$,
    $0.50$, $0.55$, $0.63$, $0.78$ and $0.98$. The line $\theta=\frac{1}{2}\Phi$
corresponding to $C_1=0.5$ separates the different operation patterns. }}
\vspace{-0cm}
\end{figure}

In Fig.~\ref{Fig:fhi_vs_Phi} obtained from Eq. (\ref{c-phase}), one sees two different XPM patterns depending on the values of 
$C_1$ and separated at the transitional point $C_1=0.5$. 
This $C_1$  value is the threshold above which the denominator
$1-C_1+C_1\cos\Phi$ on the right-hand side of Eq.~(\ref{c-phase}) 
may become zero or negative with increasing $\Phi$, 
and it is universal with respect to any pulse shape. As $C_1$ decreases or 
increases across this point, the gate performance 
characterized by the achievable $\theta$ values will undergo a transition to 
the different pattern. For $C_1<0.5$, the conditional phase $\theta$ can never 
assume values greater than $\pi/2$ no matter how large $\Phi$ is. 

Without loss of generality, in our discussion below we consider the interaction between two identical pulses in the Gaussian profile $f(z)=(\frac{1}{\sigma}\sqrt{\frac{1}{\pi}})^{\frac{1}{2}}\exp\{-\frac{z^2}{2\sigma^2}\}$. Then the boundary value $C_1=0.5$ of the two patterns in Fig. 1 corresponds to the system parameter $k_0\approx 2.5$. 
As $k_0$ increases (decreases) from this value, the coefficient $C_1$ will become smaller (larger) into the respective pattern. Eqs. (\ref{entropy}) and (\ref{fidelity}) yield the evolution of the field-mode entanglement and 
gate operation fidelity with the phase $\Phi$, which is proportional to the pulse interaction time $t$. The results 
are illustrated in Fig.~2 for various $k_0$ values. 

The expected relation between $S_L$ and $F$ manifests in the lower regime of $C_1<0.5$ in Fig. 1, where a weaker entanglement, at any fixed $\Phi$ value, is accompanied by a higher fidelity. In this regime the entanglement between the field modes totally vanishes in the limit $k_0\rightarrow \infty$. In the upper regime of $C_1>0.5$,
the field mode entanglement disappears in the limit $k_0\rightarrow 0$. The fidelity value, however, goes down to zero at this point, for any non-zero $\Phi$ value; c.f. Eq. (\ref{fidelity}) with a diverging $C_2$ at $k_0\rightarrow 0$. The non-unitary evolution due to a finite system bandwidth $\Delta\omega_s$ accounts for this phenomenon. Note that, as the consequence of non-unitary evolution, both factors $f_1$, $f_2$ in the second term of Eq. (\ref{2-p-e}) carry the same variable $z_2-vt$, thus deviating from the ideal output two-particle function $f_1(z_1-vt)f_2(z_2-vt)e^{i\theta}$ in the neighborhood of $k_0\rightarrow 0$. The XPM between two evenly distributed square pulses shows the same characteristics except for the different relations between the $C_i$ and $k_0$ values, indicating that the effects described above differ, by nature, from the inhomogeneity in pulse interactions. 

\begin{figure}[t!]
\vspace{-0cm}
\centering
\epsfig{file=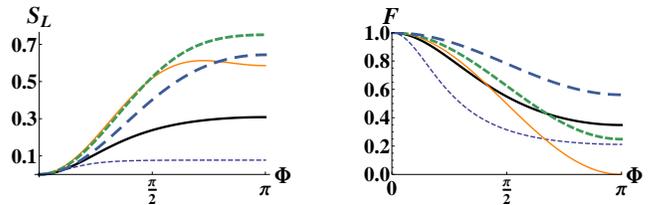,width=1.05\linewidth,clip=} 
{\vspace{-0.5cm}\caption{\label{Fig:S_LandFid_vsk0} (color online) Linear entropy $S_L$ (left) and fidelity $F$ (right) 
plotted vs $\Phi$ for various $k_0$ values: $k_0=0.5$ (thin dashed line), 
$k_0=1.0$ (thick solid line), $k_0=2.5$ (thin solid line), $k_0=5.0$ (thick short
dashed line), and $k_0=10.0$ (thick long dashed line). 
The two pulses are in the identical Gaussian profile. }}
\vspace{-0.5cm}
\end{figure}

An interesting observation from Fig. 2 is the feature that, around the transitional point $k_0\approx 2.5$ of the two operation patterns, the plots of $S_L$, e.g., the thin solid line, reaches the highest value
at some $\Phi<\pi$. Increasing $\Phi$ beyond the point leads to a continual distortion of 
the two-photon state without increasing its entanglement. Irrespective of the pulse shapes, the linear entropy plateau at the transitional point is correlated to the fidelity valley like that in Fig. 3, which goes down to $F=0$ at $k_0\approx 2.5$ and $\Phi=\pi$.

\begin{figure}[b!]
\vspace{-0.9cm}
\centering
\epsfig{file=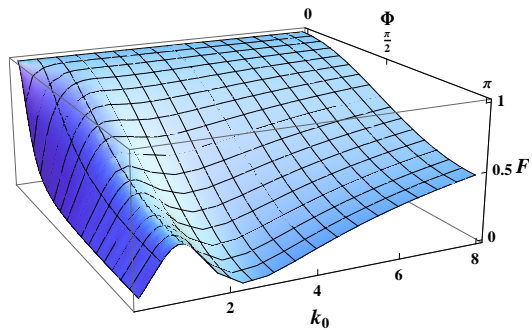, width=0.8 \linewidth,clip=} 
{\vspace{-0.5cm}\caption{(color online) Gate operation fidelity $F$ as a function of the parameters $k_0$ and $\Phi$. 
The parameter ranges are $0.1\leq k_0\leq 8$ and $0\leq \Phi\leq \pi$. It is a 3D view for the right plot of Fig. 2.}}
\vspace{-0cm}
\end{figure}

Furthermore, we comment on the notion that pulses moving with matched group velocities make large conditional phase possible \cite{l-i-00}. In the previous theoretical studies, the variables $z_1$ and $z_2$ of the two-particle function in (\ref{2-p-e}) were often mixed up with the pulse-center coordinates, and then the phase $\Phi$ would be regarded as the conditional phase $\theta$ for the reason that the overlapped pulse centers with $z_1=z_2=z$ could lead to the ideal output $f_1(z-vt)f_2(z-vt)e^{i\Phi}$ from Eq. (\ref{2-p-e}). In fact, the variable $\Delta\omega_{s}(z_1-z_2)/(2c)$ of the $\mbox{sinc}$ function in (\ref{2-p-e}) assumes any value even if the two pulses co-propagate, because $z_i$ are the field coordinates over the whole $z$ axis rather than those of the pulse centers. The conditional phase $\theta$ should be determined by Eq. (\ref{over-lap}) giving its relation with the XPM phase $\Phi$ in Fig. 1. The conditional phase value could reach $\pi$ in the upper regime of Fig. 1, where the gate operation fidelity is, however, rather low. A high fidelity is possible only in the lowest region in Fig. 1, where the state evolution is close to unitary but the peak of the conditional phase $\theta$ is vanishing. The vanishing conditional phase in the regime of near unitary evolution also exists in case of interaction between co-propagating single photon and coherent state; see \cite{bhe-11-2}. 

The trade-off between conditional phase and fidelity in XPM between co-propagating photons is also discussed recently in \cite{bana-09}, where a finite mode approximation for pulses is adopted. One problem with the finite mode approximation is its underestimation of XPM intensity---compared with the contribution from the second term in our Eq. (\ref{2-p-e}), the term arising from XPM, the discrete sum with the corresponding term in \cite{bana-09} (the second term in Eq. (22) of \cite{bana-09}) contributes much less significantly to a coefficient similar to $C_1$ in this paper, limiting its value to less than $0.5$. Then only the lower regime in our Fig. 1 can be obtained in the finite mode approach. For two ultraslow pulses the small ratio $v/c$ in the parameter $k_0$ makes the upper regime in Fig. 1 more relevant. The effect of non-unitary evolution of quantum states dominating in this regime is beyond the description by the finite mode approximation.

Next we examine the XPM between pulses colliding head-on \cite{m-f-04, f-k-05, bhe-11-1, headon-2} via a contact potential $\chi \delta(z-z')$. For two ultraslow pulses satisfying $v_r/(2c)\Delta \omega_s t  \ll 1$, there is the approximation $C(v_rt_{k-1}-v_rt_k)\approx \Delta\omega_s/(2\pi c)$ in (\ref{2-p}), so the infinite sum in Eq.~(\ref{2-p}) can be approximated by a closed form in this regime. Substituting the normalized two-particle function into 
(\ref{over-lap}), we obtain the fidelity evolution in the course of pulse interaction; see the example in Fig. 4. It shows that the fidelity value will decline once a pulse touches the other and stabilize again after they pass through each other. The stronger the interaction (indicated by the $\Phi$ values) is, the lower the fidelity will become after collision. Against the intuitive notion that an averaged interaction on pulses could generate a uniform conditional phase $\theta$, a realistic XPM between pulses of a non-zero relative velocity can be far away from the ideal process $|1\rangle_1|1\rangle_2\rightarrow e^{i\theta}|1\rangle_1|1\rangle_2$.

\begin{figure}[t!]
\vspace{-0.3cm}
\centering
\epsfig{file=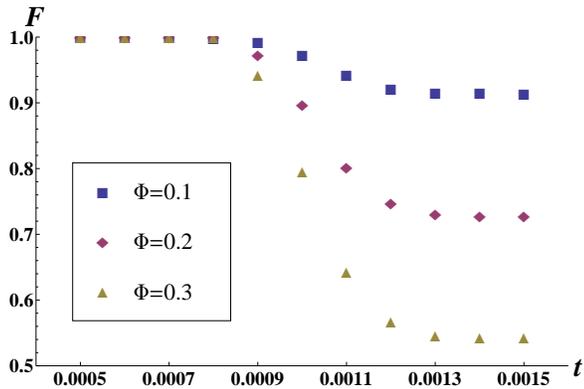, width=0.98\linewidth,clip=} 
{\vspace{-0.1cm}\caption{(color online) Fidelity evolution in head-on collision between two identical single-photon pulses 
of Gaussian profile. The two pulses are initially separated by a distance of $l=10\sigma$, where $\sigma$ is the pulse size in the medium, and run toward each other at a relative velocity $v_r=2|v_i|=10^4\sigma$ per unit time (here we adopt a scaled time unit). The pulses completely overlap at $t=0.001$. The system parameter $k_0=|v_i|/(2c)\Delta \omega_{s}/\Delta \omega_{p}$ is $0.001$. The XPM phase is defined as $\Phi=\chi \Delta \omega_s l /(\pi v_rc)$. }}
\vspace{-0.3cm}
\end{figure}

An ideal performance of phase gates based on the XPM described above exists in the limit of an infinite system bandwidth $\Delta\omega_s$. In this limit the two-particle function after two pulses completely going through each other is $f_1(z_1-vt)f_2(z_2+vt)e^{i\theta}$, where $\theta=\chi/v_r$ is a fixed value from the contact potential $\chi \delta(z-z')$. This result is also true to the XPM between single photon and coherent state \cite{bhe-11-2}. The time-dependent two-particle functions for such idealized unitary evolution under a general interaction potential $\Delta(z-z')$ take the form $f_1(z_1-v_1t)f_2(z_2-v_2t)e^{i\Phi(z_1,z_2,t)}$ \cite{f-k-05, bhe-11-1, headon-2, bhe-11-2}. The field mode entanglement exhibited by the possibly non-factorisable $\Phi(z_1,z_2,t)$ with respect to its spatial variables is therefore the main factor that determines the gate operation fidelity in a regime of approximately unitary evolution, which could be realized under the condition $\Delta \omega_s\gg \Delta\omega_p$.

\section{conclusion}

We have illustrated the effects of continuous field mode entanglement arising from pulse interaction and non-unitary evolution caused by finite system bandwidth, which drastically impair photon-photon phase gate performance. These effects induce more complexity in XPM than what was previously understood. Due to their possible existence in any device working with quantum objects of continuous degrees of freedom, the proper handling of the effects could be a major concern in quantum technology.

\appendix
\section*{Appendix}

We provide a brief derivation for the linear entropy formula in Eq. (\ref{entropy}). The elements of one reduced density matrix $\rho_1$ for a general bipartite state $\int dk\int dk'\zeta(k,k',t)\hat{a}^{\dagger}(k)\hat{b}^{\dagger}(k')|0\rangle$
can be obtained by the following \cite{s-93} (its discrete form is given in \cite{hb-08}):
\begin{eqnarray*}
\rho_1(k, k',t)=\int dq \zeta(k, q,t)\zeta^{\ast}(k', q,t)
=\frac{1}{2\pi}\int dz_1 \int dz_2&& \nonumber\\
 \times\int dz_3\big \{\psi(z_1,z_2,t)\psi^{\ast}(z_3,z_2,t)e^{-ik z_1}e^{ik'z_3}\big\}.
&&
\end{eqnarray*}
The matrix elements of $\rho^2_1$ can thus be obtained by substituting the above into $\rho^2_1(k,k',t)=\int dq \rho_1(k, q,t)\rho_1(q, k',t)$, which leads to a closed form of the linear entropy $S_L(t)=1-\int dk\rho^2_1(k,k,t)$.

\begin{acknowledgements}
We thank C.\ Simon, Y.-F.\ Chen, A.\ I.\  Lvovsky, A.~MacRae,
and P.\ M. Leung for helpful discussions. 
This work was supported by NSERC, $i$CORE, CIFAR, and General Dynamics Canada.
\end{acknowledgements}

\bigskip


\begin{thebibliography}{99}
\bibitem{ex} Our treatment as well applies to the situations with one of the pulses being a photon at the speed $c$.
\end{thebibliography}
\end{document}